\documentclass[aps,prl,twocolumn,groupedaddress,showpacs]{revtex4-1}
\usepackage{amsmath,graphicx}

\begin{document}

%\preprint{}

\title{Mechanisms of spin-dependent heat generation in spin valves}


\author{Xiao-Xue Zhang}
\author{Yao-Hui Zhu}
\email[]{yaohuizhu@gmail.com}
\author{Pei-Song He}
\author{Bao-He Li}
%\thanks{}
\affiliation{Physics Department, Beijing Technology and Business University, Beijing 100048, China}


\date{\today}

\begin{abstract}
The extra heat generation in spin transport is usually interpreted in terms of the spin relaxation. By reformulating the heat generation rate, we found alternative current-force pairs without cross effects, which enable us to interpret the product of each pair as a distinct mechanism of heat generation. The results show that the spin-dependent part of the heat generation includes two terms. One of them is proportional to the square of the spin accumulation and arises from the spin relaxation. However, the other is proportional to the square of the spin-accumulation gradient and should be attributed to another mechanism, the spin diffusion. We illustrated the characteristics of the two mechanisms in a typical spin valve with a finite nonmagnetic spacer layer.

%The heat due to the spin relaxation is proportional to the square of the spin accumulation, whereas the heat due to the spin diffusion is proportional to the square of the spin-accumulation gradient.

%The two mechanisms have equal contributions in semi-infinite layers, such as the ferromagnetic layers of a spin valve. However, their relative magnitude varies with the layer thickness in a finite layer, such as the nonmagnetic layer of the spin valve. When the NM-layer thickness is much smaller than its spin-diffusion length, its spin-dependent heat generation is dominated by the spin relaxation in the antiparallel configuration, and by the spin diffusion in the parallel configuration.

%The heat generated via the two mechanisms can be expressed formally as the Joule heat of two effective resistances in an effective circuit.
\end{abstract}

\pacs{75.40.Gb, 85.75.-d, 75.76.+j, 76.60.Es, 91.45.Rg}
% spin diffusion: 75.40.Gb
% magnetic devices - spin polarized transport devices: 85.75.-d
% spin transport effects: 75.76.+j
% spin-lattice relaxation: 76.60.Es
% spin-polarized transport processes: 72.25.-b
% heat generation and transport: 91.45.Rg

%\keywords{}

\maketitle


Heat generation plays an important role in spintronic devices.~\cite{Bauer2012,Adachi2013} Experimental studies have demonstrated various spin-dependent heating effects.~\cite{Gravier06,E:resis2010,Slachter2011,E:valve2012,Peltier2012,E:T-O2014,Marun16}
Recent theoretical investigations have also shown that there is still dissipation even if a pure spin current is present.~\cite{T:Boltz2011,Wegrowe2011,Slachter2011Nov,T:pump2014,Juarez16} The extra heat associated with the spin degree of freedom is usually attributed to a single mechanism, the spin relaxation (or the spin-flip scattering).~\cite{T:Boltz2011} Previous works did not pay much attention to the distinctions and characteristics of the various mechanisms of heat generation. Here, by reformulating the heat generation rate, we show that the spin-dependent part of the heat generation actually arises from two mechanisms: the spin relaxation and the spin diffusion. Moreover, it is also worthwhile to compare the two mechanisms in a basic spintronic device, the spin valve.

We consider a typical spin valve composed of two semi-infinite ferromagnetic (FM) layers and a finite nonmagnetic (NM) spacer layer. A DC of density $J$ flows in the positive $z$-direction that is perpendicular to the layer plane.~\cite{Fert1993} In the stationary state, the heat generation rate $\sigma_\mathrm{heat}$ in an FM or NM layer can be written as
\begin{equation}\label{eq:heat2}
\sigma_\mathrm{heat}=\frac{J}{e}\frac{\partial\bar{\mu}}{\partial z}
+\frac{J_\mathrm{spin}}{e}\frac{\partial\Delta{\mu}}{\partial z}
+\frac{4(\Delta\mu)^2}{e^2}G_\mathrm{mix}
\end{equation}
%+\frac{\partial{J}_\mathrm{spin}}{\partial{z}}\frac{\Delta\mu}{e}
where $\bar\mu=(\bar{\mu}_{+}+\bar{\mu}_{-})/2$ is the average electrochemical potential and $-e$ the charge of an electron. This equation can be derived by rewriting Eq.~(6) of Ref.~\cite{T:Boltz2011} in terms of the total current $J=J_{+}+J_{-}$ and the spin current $J_\mathrm{spin}=J_{+}-J_{-}$. Here $\bar\mu_{+}$ ($\bar\mu_{-}$) and $J_{+}$ ($J_{-}$) are the electrochemical potential and the current density in the spin-up (down) channel, respectively. Moreover, $\Delta\mu=(\bar\mu_{+}-\bar\mu_{-})/2$ describes the spin accumulation, and $G_\mathrm{mix}$ is the spin-flip rate, which is defined by Eq.~(5) of Ref.~\cite{T:Boltz2011}.

The meaning of Eq.~(\ref{eq:heat2}) can be interpreted as follows. The last term stands for the heat generated by the spin-flip scattering and we denote it by $\sigma_\mathrm{heat}^\mathrm{sf}$. This term can be written in a form with transparent interpretation
\begin{equation}\label{heatsf1}
\sigma_\mathrm{heat}^\mathrm{sf}=\frac{N_s(2\Delta\mu)^2}{\tau_\mathrm{sf}}
\end{equation}
by using $G_\mathrm{mix}=e^2N_s/\tau_\mathrm{sf}$, where $N_s$ is the density of states for spin $s$ and $\tau_\mathrm{sf}$ the spin-flip relaxation time at the Fermi energy.~\cite{T:Boltz2011} To avoid the calculation of $N_s$, we first derive the equation
\begin{equation}\label{jspinderivative}
\frac{\partial{J}_\mathrm{spin}}{\partial{z}}
=\frac{4G_\mathrm{mix}\Delta\mu}{e}=\frac{4eN_s\Delta\mu}{\tau_\mathrm{sf}}
%=\frac{\Delta\mu}{er_\mathrm{F(N)}l_\mathrm{sf}^\mathrm{F(N)}}
\end{equation}
from Eq.~(4a) of Ref.~\cite{T:Boltz2011}. Then we can rewrite $\sigma_\mathrm{heat}^\mathrm{sf}$ as
\begin{equation}\label{heatsf}
\sigma_\mathrm{heat}^\mathrm{sf}=\frac{\partial{J}_\mathrm{spin}}{\partial{z}}\frac{\Delta\mu}{e}
%=\frac{\tau_\mathrm{sf}}{4N_s}\left(\frac{1}{e}\frac{\partial{J}_\mathrm{spin}}{\partial{z}}\right)^2
\end{equation}
where $J_\mathrm{spin}$ and $\Delta\mu$ can be derived by using the Valet-Fert theory.~\cite{Fert1993} As for the first term of $\sigma_\mathrm{heat}$, we need Eq.~(13) of Ref.~\cite{Fert1993}
\begin{equation}
J_\pm=\sigma_\pm\left(F\pm{F}_\mathrm{sd}\right)\label{eq:conti:Ohm's law2}
\end{equation}
where the field $F$ is defined as $F=(1/e)\left(\partial\bar\mu/\partial{z}\right)$. We have also introduced the generalized force
\begin{equation}
F_\mathrm{sd}=\frac{1}{e}\frac{\partial\Delta\mu}{\partial{z}}\label{sdforce}
\end{equation}
which results from the spin diffusion and contains only exponential terms. The conductivity $\sigma_{\uparrow(\downarrow)}$ can be written as $1/\sigma_{\uparrow(\downarrow)}=\rho_{\uparrow(\downarrow)}=2\rho^\ast_\mathrm{F}[1-(+)\beta]$ in an FM layer, where $\uparrow$ ($\downarrow$) denotes majority (minority) spin direction.~\cite{Fert1993} The bulk spin asymmetry coefficient $\beta$ is positive in FM layers. The equation for $\sigma_{\uparrow(\downarrow)}$ can also be applied to the NM layer by simply setting $\beta=0$ and replacing $\rho_\mathrm{F}^\ast$ by $\rho_\mathrm{N}^\ast$. Summing the ``$\pm$'' components of Eq.~\eqref{eq:conti:Ohm's law2} yields
\begin{equation}
F=E_0\pm\beta{F}_\mathrm{sd}\label{eq:meanfield}
\end{equation}
where $E_0=(1-\beta^2)\rho^\ast{J}$ is the constant unperturbed electric field and ``$+$'' (``$-$'') corresponds to ``up'' (``down'') magnetization.~\cite{Fert1993} In the NM layer, we have $F=E_0^\mathrm{N}=\rho_\mathrm{N}^\ast{J}$. Then the first term of $\sigma_\mathrm{heat}$ becomes $JE_0^\mathrm{N}=\rho_\mathrm{N}^\ast{J}^2$ and stands for the nominal Joule heat, which does not depend on the spin accumulation. However, in the FM layers, this term contains the nominal Joule heat $JE_0^\mathrm{F}=(1-\beta^2)\rho_\mathrm{F}^\ast{J}^2$ as well as a spin-dependent term, $\pm{J}\beta{F}_\mathrm{sd}$, which is difficult to interpret. A similar difficulty also exists in the second term of $\sigma_\mathrm{heat}$. One can see this by subtracting the ``$\pm$'' components of Eq.~(\ref{eq:conti:Ohm's law2})
\begin{equation}\label{spincurrent}
J_\mathrm{spin}=J_\mathrm{spin}^\mathrm{bulk}+J_\mathrm{spin}^\mathrm{exp}
\end{equation}
where the total spin current, $J_\mathrm{spin}$, has been divided into a bulk term and an exponential term. The bulk term is defined as $J_\mathrm{spin}^\mathrm{bulk}=\mp\beta{J}$, where ``$-$'' (``$+$'') corresponds to ``up'' (``down'') magnetization. The exponential term can be written as
\begin{equation}
J_\mathrm{spin}^\mathrm{exp}=\frac{1}{\rho^\ast}F_\mathrm{sd}
=\left(1-\beta^2\right)\sigma{F}_\mathrm{sd}\label{spinexp}
\end{equation}
where $\sigma=\sigma_++\sigma_-$ is the total conductivity. In the NM layers, we have $J_\mathrm{spin}=J_\mathrm{spin}^\mathrm{exp}$ and then the second term of $\sigma_\mathrm{heat}$ becomes $J_\mathrm{spin}^\mathrm{exp}F_\mathrm{sd}$. The previous works did not pay much attention to this term and we will interpret it below. In the FM layers, $J_\mathrm{spin}^\mathrm{bulk}$ is not zero and then the second term of $\sigma_\mathrm{heat}$ also contains another term, $J_\mathrm{spin}^\mathrm{bulk}F_\mathrm{sd}$, which is also difficult to interpret.

To continue our study on the mechanisms of heat generation, we have to overcome the difficulties in the FM layers. These difficulties result from the cross effects between the two generalized current-force pairs chosen in Eq.~(\ref{eq:heat2}): ($J$, $F$) and ($J_\mathrm{spin}$, $F_\mathrm{sd}$).~\cite{T:Boltz2011,Kondepudi} The cross effects can be better demonstrated by the matrix equation
\begin{equation}\label{matrix1}
\left(
\begin{array}{c}
J\\
J_\mathrm{spin}
\end{array}\right)
=\left(\begin{array}{cc}
\sigma &\quad \sigma_+-\sigma_-\\
\sigma_+-\sigma_- &\quad \sigma
\end{array}\right)
\left(
\begin{array}{c}
F\\
F_\mathrm{sd}
\end{array}\right)
\end{equation}
which can be derived by summing and subtracting the ``$\pm$'' components of Eq.~\eqref{eq:conti:Ohm's law2}, respectively. The off-diagonal elements of the $2\times2$ matrix are zero in the NM layer, whereas they are not in the FM layers and this leads to the cross effects.

%Then, the cross effects lead to cross terms in the product of the current-force pairs, $JF$ and $J_\mathrm{spin}F_\mathrm{sd}$, which are just the first two terms of $\sigma_\mathrm{heat}$ in Eq.~(\ref{eq:heat2}).

%The cross terms here are not $\pm(J\beta/e)(\partial\Delta\mu/\partial{z})$ and $(J_\mathrm{spin}^\mathrm{bulk}/e)(\partial\Delta\mu/\partial{z})$

%rewriting Eq.~(\ref{eq:heat2}) with

We can avoid the cross effects by choosing appropriate current-force pairs. Substituting Eqs.~(\ref{eq:meanfield}) and (\ref{spincurrent}) into Eq.~(\ref{eq:heat2}), we can rewrite the first two terms of $\sigma_\mathrm{heat}$ as
\begin{equation}\label{eq:heat3}
\sigma_\mathrm{heat}^\mathrm{nom}+\sigma_\mathrm{heat}^\mathrm{sd}
=JE_0+J_\mathrm{spin}^\mathrm{exp}F_\mathrm{sd}
\end{equation}
where we have introduced $\sigma_\mathrm{heat}^\mathrm{nom}=JE_0$ and
\begin{equation}\label{heatsc}
\sigma_\mathrm{heat}^\mathrm{sd}=J_\mathrm{spin}^\mathrm{exp}F_\mathrm{sd}
=\frac{1}{\rho^\ast}\left(F_\mathrm{sd}\right)^2
\end{equation}
Then we choose ($J$, $E_0$) and ($J_\mathrm{spin}^\mathrm{exp}$, $F_\mathrm{sd}$) as the new current-force pairs, and they satisfy the equation
\begin{equation}\label{matrix2}
\left(
\begin{array}{c}
J\\
J_\mathrm{spin}^\mathrm{exp}
\end{array}\right)
=\left[\begin{array}{cc}
\sigma &\quad 0\\
0 &\quad (1-\beta^2)\sigma
\end{array}\right]
\left(
\begin{array}{c}
E_0\\
F_\mathrm{sd}
\end{array}\right)
\end{equation}
which is the matrix form of $E_0=(1-\beta^2)\rho^\ast{J}=J/\sigma$ and Eq.~(\ref{spinexp}). Now the off-diagonal elements are zero in both FM and NM layers, and the cross effects disappear. We interpret $\sigma_\mathrm{heat}^\mathrm{sd}$ in Eq.~(\ref{heatsc}) as the heat generation due to the spin diffusion. The reason is that, the gradient of spin accumulation ($\propto{F}_\mathrm{sd}$) drives spins from a position to another one with lower chemical potential via the diffusion process in the same spin channel. The loss in the chemical potential of the spins leads to heat generation. This can also be understood by writing this term as the sum of the contributions from the two spin channels. The spin diffusion is especially important for the spin-dependent heat generation at interfaces, because the spin relaxation is usually negligible and the spin diffusion is the dominant mechanism here.~\cite{Fert1993} We stress that $\sigma_\mathrm{heat}^\mathrm{sd}$ should not be interpreted as the heat generation due to the spin-flip scattering together with $\sigma_\mathrm{heat}^\mathrm{sf}$.

Then we can reformulate the heat generation rate as
\begin{equation}\label{eq:heatf}
%\begin{split}
\sigma_\mathrm{heat}=\sigma_\mathrm{heat}^\mathrm{nom}
+\sigma_\mathrm{heat}^\mathrm{sd}
+\sigma_\mathrm{heat}^\mathrm{sf}\\
%&=(1-\beta^2)\rho_\mathrm{F(N)}^\ast{J}^2
%+\rho_\mathrm{F(N)}^\ast\left(J_\mathrm{spin}^\mathrm{exp}\right)^2
%+\rho_\mathrm{F(N)}^\ast\left[\frac{\Delta\mu}{er_\mathrm{F(N)}}\right]^2
%\end{split}
\end{equation}
where $\sigma_\mathrm{heat}^\mathrm{nom}$ exists no matter if the spin accumulation is present or not. However, $\sigma_\mathrm{heat}^\mathrm{sf}$ is proportional to the square of the spin accumulation as shown by Eq.~\eqref{heatsf1}. Meanwhile, $\sigma_\mathrm{heat}^\mathrm{sd}$ is proportional to the square of the spin-accumulation \emph{gradient} as shown by Eqs.~(\ref{sdforce}) and (\ref{heatsc}).

Now we are ready to compare $\sigma_\mathrm{heat}^\mathrm{sd}$ and $\sigma_\mathrm{heat}^\mathrm{sf}$ in two typical cases: a semi-infinite layer and a finite layer. The spin valve contains both types of layers: the semi-infinite FM layers and the NM layer with a finite thickness. We place the origin of the $z$-axis at the center of the NM layer. The left and right FM/NM interfaces are located at $z=-d$ and $z=d$, respectively. The two FM layers are made of the same material with collinear magnetization. Without loss of generality, the magnetization direction of the left FM layer is fixed and set to be ``up''. The right FM layer is assumed to have ``down''  and ``up'' magnetization for antiparallel (AP) and parallel (P) alignments, respectively. The interface resistance is neglected since its spin-dependent heat generation is dominated by the spin diffusion and no comparison is necessary.

%The various quantities in Eq.~\eqref{eq:heatf} can be derived according to the Valet-Fert theory.~\cite{Fert1993}

In the two FM layers, both $\sigma_\mathrm{heat}^\mathrm{sf,AP(P)}$ and $\sigma_\mathrm{heat}^\mathrm{sd,AP(P)}$ are even functions as shown by Eqs.~\eqref{heatsf1} and (\ref{heatsc}). Then we need consider only the left FM layer ($z<-d$) without loss of generality. Using Eqs.~\eqref{heatsf} and (\ref{heatsc}), we have
\begin{equation}\label{heatfm1}
\sigma_\mathrm{heat}^\mathrm{sd,AP(P)}=\sigma_\mathrm{heat}^\mathrm{sf,AP(P)}
=\frac{\Sigma_\mathrm{heat}^\mathrm{AP(P),F}}{l_\mathrm{sf}^\mathrm{F}}\exp
\left[\frac{2(z+d)}{l_\mathrm{sf}^\mathrm{F}}\right]
\end{equation}
where $\Sigma_\mathrm{heat}^\mathrm{AP(P),F}=r_\mathrm{F}\left[\alpha_\mathrm{F}^\mathrm{AP(P)}J\right]^2$ and $r_\mathrm{F}=\rho_\mathrm{F}^\ast{l}_\mathrm{sf}^\mathrm{F}$. We have introduced the dimensionless parameter
\begin{equation}
\alpha_\mathrm{F}^\mathrm{AP(P)}=\frac{\beta r_\mathrm{N}^\mathrm{AP(P)}}{r_\mathrm{F}+r_\mathrm{N}^\mathrm{AP(P)}}\label{alphafnoir}
\end{equation}
where $r_\mathrm{N}^\mathrm{AP}$ ($r_\mathrm{N}^\mathrm{P}$) is defined as $r_\mathrm{N}^\mathrm{AP}=r_\mathrm{N}\coth\xi$ ($r_\mathrm{N}^\mathrm{P}=r_\mathrm{N}\tanh\xi$). Here we also have $r_\mathrm{N}=\rho_\mathrm{N}^\ast{l}_\mathrm{sf}^\mathrm{N}$ and $\xi=d/l_\mathrm{sf}^\mathrm{N}$. Equation~(\ref{heatfm1}) shows that the spin diffusion leads to the same heat generation at any position as the spin relaxation in \emph{semi-infinite} layers. Their integrals from $-\infty$ to $-d$ are also equal to each other
\begin{align}
&\Sigma_\mathrm{heat}^\mathrm{sd,AP(P),F}=\int_{-\infty}^{-d}
\sigma_\mathrm{heat}^\mathrm{sd,AP(P)}dz
=r^\mathrm{sd}_\mathrm{F}\left[J_\mathrm{F,sf}^\mathrm{AP(P)}\right]^2\label{totalh1}\\
&\Sigma_\mathrm{heat}^\mathrm{sf,AP(P),F}=\int_{-\infty}^{-d}
\sigma_\mathrm{heat}^\mathrm{sf,AP(P)}dz
=r^\mathrm{sf}_\mathrm{F}\left[J_\mathrm{F,sf}^\mathrm{AP(P)}\right]^2\label{totalh2}
\end{align}
where we have $r^\mathrm{sd}_\mathrm{F}=r^\mathrm{sf}_\mathrm{F}=2r_\mathrm{F}$ and $J_\mathrm{F,sf}^\mathrm{AP(P)}=\alpha_\mathrm{F}^\mathrm{AP(P)}J/2$. One can verify that $\Sigma_\mathrm{heat}^\mathrm{AP(P),F}=\Sigma_\mathrm{heat}^\mathrm{sd,AP(P),F}+\Sigma_\mathrm{heat}^\mathrm{sf,AP(P),F}$ is the spin-dependent part of the total heat generation in the left FM layer.

In the NM layer, the heat generation has different characteristics for AP and P alignments. For the AP alignment, the two spin-dependent terms can be written as
\begin{align}
\sigma_\mathrm{heat}^\mathrm{sd,AP}&=
\frac{2\Sigma_\mathrm{heat}^\mathrm{AP,N}}{l_\mathrm{sf}^\mathrm{N}\sinh(2\xi)}
\sinh^2\left(\frac{z}{l_\mathrm{sf}^\mathrm{N}}\right)\label{heatnmapsc}\\
\sigma_\mathrm{heat}^\mathrm{sf,AP}&=
\frac{2\Sigma_\mathrm{heat}^\mathrm{AP,N}}{l_\mathrm{sf}^\mathrm{N}\sinh(2\xi)}
\cosh^2\left(\frac{z}{l_\mathrm{sf}^\mathrm{N}}\right)\label{heatnmapsf}
\end{align}
where $\Sigma_\mathrm{heat}^\mathrm{AP,N}=r_\mathrm{N}^\mathrm{AP}\left(\alpha_\mathrm{N}^\mathrm{AP}J\right)^2$.
For the P alignment, we have
\begin{align}
\sigma_\mathrm{heat}^\mathrm{sd,P}&=
\frac{2\Sigma_\mathrm{heat}^\mathrm{P,N}}{l_\mathrm{sf}^\mathrm{N}
\sinh(2\xi)}\cosh^2\left(\frac{z}{l_\mathrm{sf}^\mathrm{N}}\right)\label{heatnmpsc}\\
\sigma_\mathrm{heat}^\mathrm{sf,P}&=
\frac{2\Sigma_\mathrm{heat}^\mathrm{P,N}}{l_\mathrm{sf}^\mathrm{N}\sinh(2\xi)}
\sinh^2\left(\frac{z}{l_\mathrm{sf}^\mathrm{N}}\right)\label{heatnmpsf}
\end{align}
where $\Sigma_\mathrm{heat}^\mathrm{P,N}=r_\mathrm{N}^\mathrm{P}\left(\alpha_\mathrm{N}^\mathrm{P}J\right)^2$.
The dimensionless parameter $\alpha_\mathrm{N}^\mathrm{AP(P)}$ is defined as
\begin{equation}
\alpha_\mathrm{N}^\mathrm{AP(P)}=\frac{\beta r_\mathrm{F}}{r_\mathrm{F}+r_\mathrm{N}^\mathrm{AP(P)}}\label{alphannoir}
\end{equation}
which satisfies the identity $\alpha_\mathrm{F}^\mathrm{AP(P)}+\alpha_\mathrm{N}^\mathrm{AP(P)}=\beta$.

\begin{figure}
\includegraphics[width=0.48\textwidth]{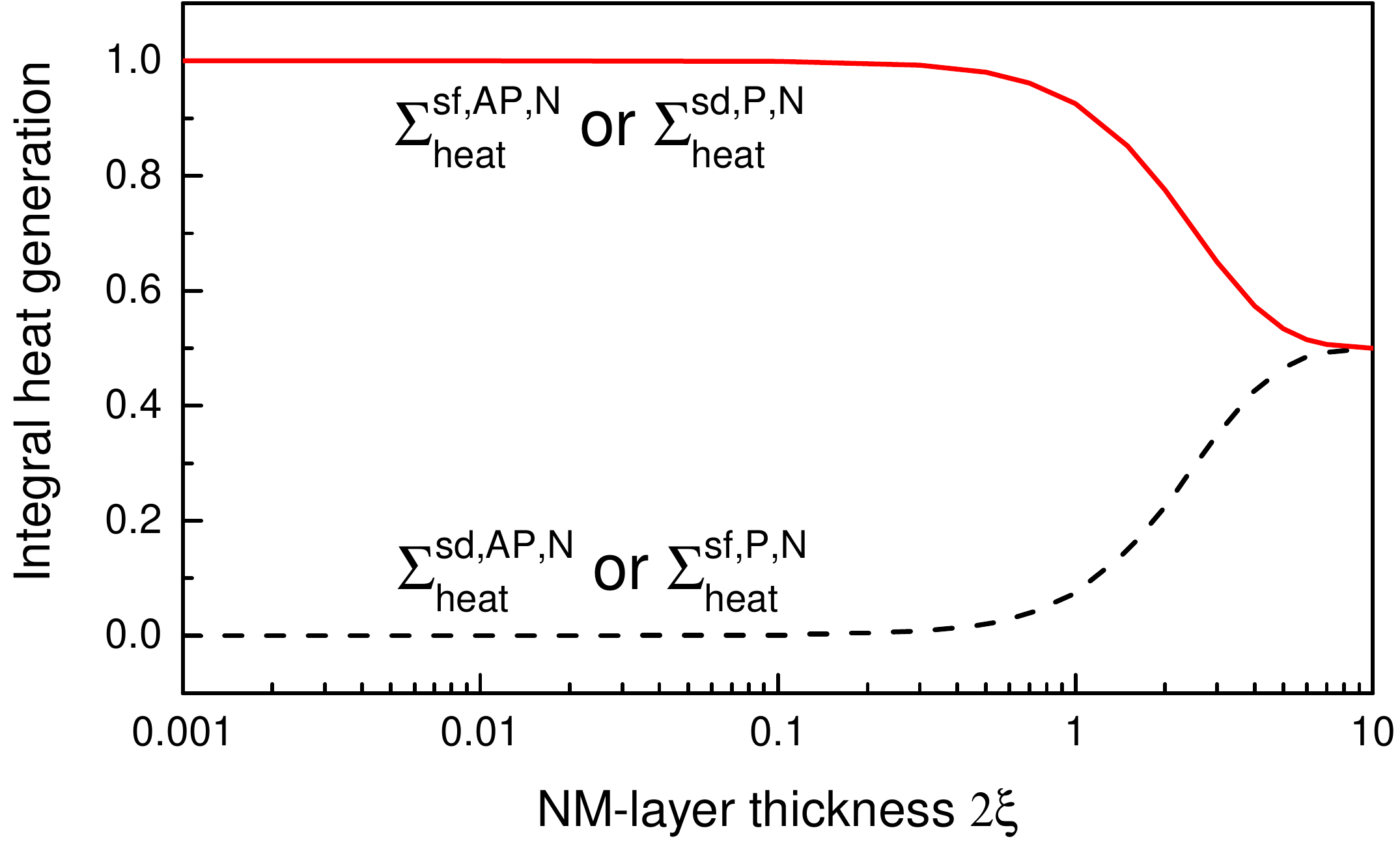}
\caption{\label{fig-relative}Integral heat generation rate (in unit of $\Sigma_\mathrm{heat}^\mathrm{AP,N}$ and $\Sigma_\mathrm{heat}^\mathrm{P,N}$ for AP and P alignments, respectively) in half of the NM layer as a function of the NM-layer thickness (in unit of $l_\mathrm{sf}^\mathrm{N}$). The red-solid curve corresponds to the integral heat generation due to the spin relaxation (diffusion) in the AP (P) configuration, and the black-dashed curve stands for that due to the spin diffusion (relaxation) in the AP (P) configuration.}
\end{figure}

For the NM layer, it is more meaningful to compare the integral heat generation due to the two mechanisms. Without loss of generality, we need consider only the left half of the NM layer since both $\sigma_\mathrm{heat}^\mathrm{sd,AP(P)}$ and $\sigma_\mathrm{heat}^\mathrm{sf,AP(P)}$ are even functions. For the AP alignment, we have
\begin{align}
\Sigma_\mathrm{heat}^\mathrm{sd,AP,N}&=\int_{-d}^{0}\sigma_\mathrm{heat}^\mathrm{sd,AP}dz
=r_\mathrm{N}^\mathrm{sd,AP}\left(J_\mathrm{sf,N}^\mathrm{AP}\right)^2\label{sigmanapsc}\\
\Sigma_\mathrm{heat}^\mathrm{sf,AP,N}&=\int_{-d}^{0}\sigma_\mathrm{heat}^\mathrm{sf,AP}dz
=r_\mathrm{N}^\mathrm{sf,AP}\left(J_\mathrm{sf,N}^\mathrm{AP}\right)^2\label{sigmanapsf}
\end{align}
where $r_\mathrm{N}^\mathrm{sd,AP}=2\left(1-\eta\right)r_\mathrm{N}^\mathrm{AP}$, $r_\mathrm{N}^\mathrm{sf,AP}=2\left(1+\eta\right)r_\mathrm{N}^\mathrm{AP}$, and $J_\mathrm{sf,N}^\mathrm{AP}=\alpha_\mathrm{N}^\mathrm{AP}J/2$. One can verify that $\Sigma_\mathrm{heat}^\mathrm{AP,N}=\Sigma_\mathrm{heat}^\mathrm{sd,AP,N}+\Sigma_\mathrm{heat}^\mathrm{sf,AP,N}$ is the spin-dependent part of the total heat generation in the left half of the NM layer. The dimensionless parameter $\eta=2\xi/\sinh(2\xi)$ describes the asymmetry between $\Sigma_\mathrm{heat}^\mathrm{sd,AP,N}$ and $\Sigma_\mathrm{heat}^\mathrm{sf,AP,N}$. This becomes more obvious if one looks at their relative magnitude $(\Sigma_\mathrm{heat}^\mathrm{sf,AP,N}-\Sigma_\mathrm{heat}^\mathrm{sd,AP,N})
/\Sigma_\mathrm{heat}^\mathrm{AP,N}=\eta$,
which decreases from $1$ to $0$ as $\xi$ varies from $0$ to $\infty$. Figure~\ref{fig-relative} shows their variation with the NM layer thickness $2\xi$ (in unit of $l_\mathrm{sf}^\mathrm{N}$). In the regime of $2\xi=2d/l_\mathrm{sf}^\mathrm{N}\ll{1}$, which is practical for application, $\Sigma_\mathrm{heat}^\mathrm{sd,AP,N}/\Sigma_\mathrm{heat}^\mathrm{AP,N}$ and $\Sigma_\mathrm{heat}^\mathrm{sf,AP,N}/\Sigma_\mathrm{heat}^\mathrm{AP,N}$ approach $0$ and $1$, respectively. Therefore, the spin-dependent heat generation of the NM layer is dominated by the spin relaxation for the AP alignment. This behavior can be interpreted as follows. We have $z/l_\mathrm{sf}^\mathrm{N}\ll{1}$ in the NM layer ($-d<z<d$) if $2d/l_\mathrm{sf}^\mathrm{N}\ll{1}$. This allows us to expand the spin accumulation $\Delta\mu=er_\mathrm{N}^\mathrm{AP}\alpha_\mathrm{N}^\mathrm{AP}J
\cosh(z/l_\mathrm{sf}^\mathrm{N})/\cosh\xi$ in terms of $z/l_\mathrm{sf}^\mathrm{N}$ and keep up to the first-order term. The result is a term independent of position since the first-order term is absent. Then the gradient of $\Delta\mu$ becomes zero and the spin diffusion is suppressed. Therefore, $\sigma_\mathrm{heat}^\mathrm{sd,AP}$ approaches zero in this regime according to Eq.~\eqref{heatsc}.

%Then $J_\mathrm{spin}^\mathrm{exp}$ approaches zero because it is proportional to the gradient of $\Delta\mu$ as shown by Eq.~\eqref{spinexp}.

For the P alignment, the integral heat generation due to the two mechanisms can be written similarly as
\begin{align}
\Sigma_\mathrm{heat}^\mathrm{sd,P,N}&=\int_{-d}^{0}\sigma_\mathrm{heat}^\mathrm{sd,P}dz
=r_\mathrm{N}^\mathrm{sd,P}\left(J_\mathrm{sf,N}^\mathrm{P}\right)^2\label{sigmanpsc}\\
\Sigma_\mathrm{heat}^\mathrm{sf,P,N}&=\int_{-d}^{0}\sigma_\mathrm{heat}^\mathrm{sf,P}dz
=r_\mathrm{N}^\mathrm{sf,P}\left(J_\mathrm{sf,N}^\mathrm{P}\right)^2\label{sigmanpsf}
\end{align}
where $r_\mathrm{N}^\mathrm{sd,P}=2\left(1+\eta\right)r_\mathrm{N}^\mathrm{P}$, $r_\mathrm{N}^\mathrm{sf,P}=2\left(1-\eta\right)r_\mathrm{N}^\mathrm{P}$, and $J_\mathrm{sf,N}^\mathrm{P}=\alpha_\mathrm{N}^\mathrm{P}J/2$. Similarly, $\Sigma_\mathrm{heat}^\mathrm{P,N}=\Sigma_\mathrm{heat}^\mathrm{sd,P,N}+\Sigma_\mathrm{heat}^\mathrm{sf,P,N}$ is the spin-dependent part of the total heat generation in the left half of the NM layer. However, the relative magnitude of $\Sigma_\mathrm{heat}^\mathrm{sd,P,N}$ and $\Sigma_\mathrm{heat}^\mathrm{sf,P,N}$ is switched in comparison to the AP alignment and the spin diffusion becomes dominant in the regime $2d/l_\mathrm{sf}^\mathrm{N}\ll{1}$ (see Fig.~\ref{fig-relative}). This can be interpreted in a similar way to the AP alignment. The lowest-order term in the expansion of the spin current $J_\mathrm{spin}=-\alpha_\mathrm{N}^\mathrm{P}J\cosh(z/l_\mathrm{sf}^\mathrm{N})/\cosh\xi$ is independent on position. Because $\Delta\mu$ is proportional to the gradient of $J_\mathrm{spin}$ as shown by Eq.~(\ref{jspinderivative}), it becomes zero and then the spin relaxation is suppressed. Therefore, $\sigma_\mathrm{heat}^\mathrm{sf,P}$ approaches zero in this regime according to Eq.~\eqref{heatsf1} or (\ref{heatsf}). In the opposite regime, $2\xi=2d/l_\mathrm{sf}^\mathrm{N}\gg{1}$, the integral heat generation $\Sigma_\mathrm{heat}^\mathrm{sd,N}$ and $\Sigma_\mathrm{heat}^\mathrm{sf,N}$ approach the same value for both P and AP alignments as shown in Fig.~\ref{fig-relative}. The semi-infinite case (the FM layers) is recovered in this limit.

%Thus the spin diffusion has the same contribution as the spin relaxation when the NM-layer thickness becomes much larger than the spin-diffusion length.
%Then $\Delta\mu$ approaches zero because it is proportional to the gradient of $J_\mathrm{spin}$ as shown by Eq. \eqref{jspinderivative}.

\begin{figure}
\includegraphics[width=0.48\textwidth]{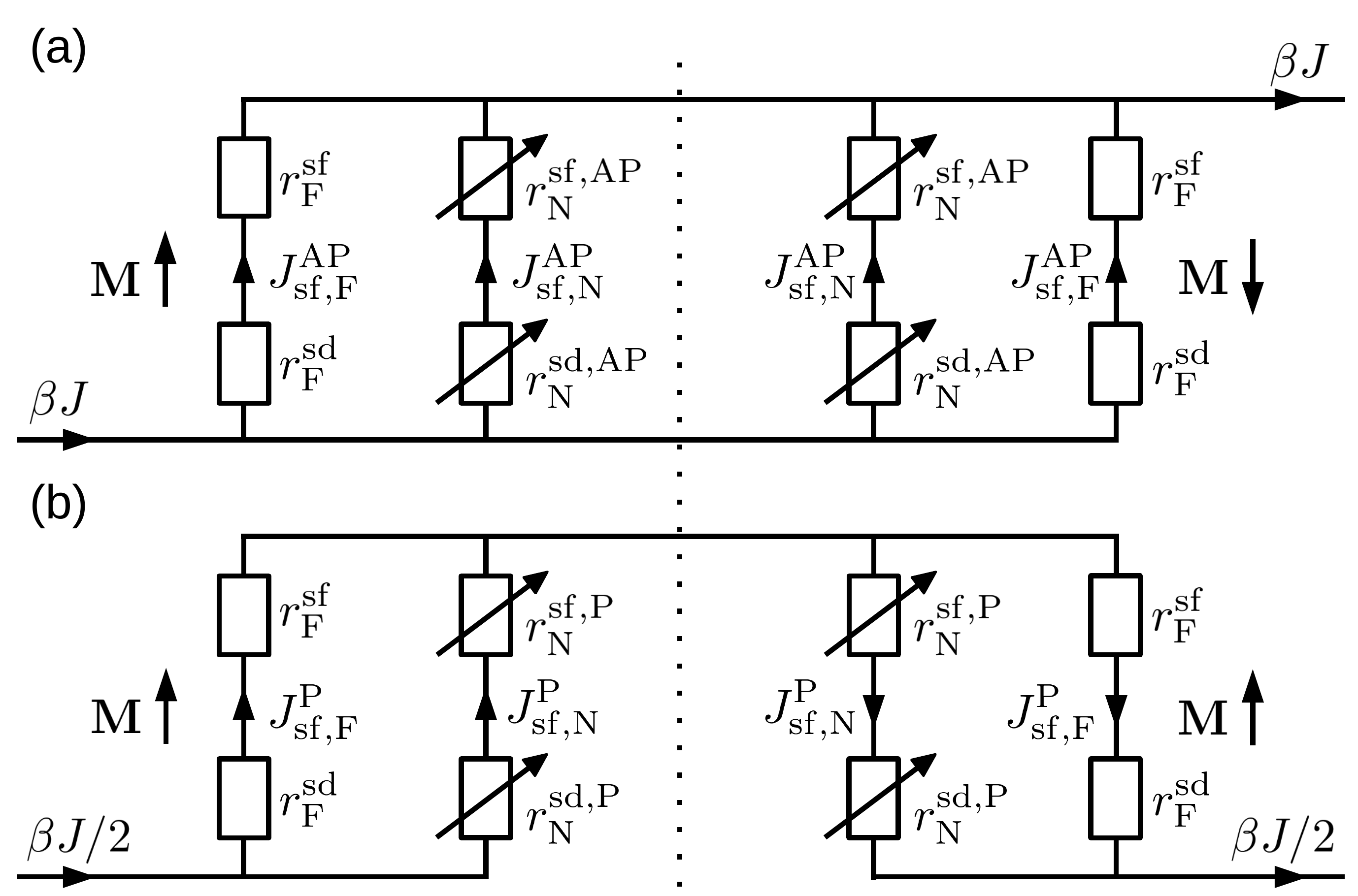}
\caption{\label{fig:circuit}Effective circuits for the spin-dependent heat generation in the spin valve. The vertical dotted line indicates the center of the spin valve and $\mathbf{M}$ together with an arrow ($\uparrow$ or $\downarrow$) shows the magnetization direction of an FM layer. (a) In the AP configuration, the resistances $r_\mathrm{F}^\mathrm{sd}$ and $r_\mathrm{F}^\mathrm{sf}$ are defined in Eqs.~\eqref{totalh1} and (\ref{totalh2}), respectively. The variable resistances $r_\mathrm{N}^\mathrm{sd,AP}$ and $r_\mathrm{N}^\mathrm{sf,AP}$ are defined in Eqs.~(\ref{sigmanapsc}) and (\ref{sigmanapsf}), respectively. (b) In the P configuration, $r_\mathrm{F}^\mathrm{sd}$ and $r_\mathrm{F}^\mathrm{sf}$ are the same as those in (a), whereas $r_\mathrm{N}^\mathrm{sd,P}$ and $r_\mathrm{N}^\mathrm{sf,P}$ are defined in (\ref{sigmanpsc}) and (\ref{sigmanpsf}), respectively.}
\end{figure}

%as $2(1-\eta)r_\mathrm{N}^\mathrm{AP}$ and $2(1+\eta)r_\mathrm{N}^\mathrm{AP}$
%as $2(1+\eta)r_\mathrm{N}^\mathrm{P}$ and $2(1-\eta)r_\mathrm{N}^\mathrm{P}$

The spin-dependent heat generation in the spin valve can also be interpreted by applying Joule's law to the effective circuits shown in Fig.~\ref{fig:circuit}. In the AP configuration shown by Fig.~\ref{fig:circuit}(a), the current of density $\beta{J}/2$ flows from the spin-down to the spin-up channel in each half of the spin valve. It models the electron-number current (spin flux in Ref.~\cite{Wegrowe2011}) flowing inversely due to the spin-flip scattering in the FM and NM layers. The current density $J_\mathrm{sf,F}^\mathrm{AP}=\alpha_\mathrm{F}^\mathrm{AP}J/2$ and $J_\mathrm{sf,N}^\mathrm{AP}=\alpha_\mathrm{N}^\mathrm{AP}J/2$ can also be derived from the circuit by using $J_\mathrm{sf,F}^\mathrm{AP}+J_\mathrm{sf,N}^\mathrm{AP}=\beta{J}/2$, which is a transformation of the identity $\alpha_\mathrm{F}^\mathrm{AP}+\alpha_\mathrm{N}^\mathrm{AP}=\beta$. The heat generation due to the spin relaxation in either FM (half of the NM) layer is modeled by the Joule heating of the resistance $r_\mathrm{F}^\mathrm{sf}$ ($r^\mathrm{sf,AP}_\mathrm{N}$). Similarly, the heat generation due to the spin diffusion is modeled by the Joule heat of the resistances $r_\mathrm{F}^\mathrm{sd}$ and $r^\mathrm{sd,AP}_\mathrm{N}$. One can recover Eqs.~\eqref{totalh1}, (\ref{totalh2}), \eqref{sigmanapsc}, and \eqref{sigmanapsf} according to Joule's law.

In the P configuration shown by Fig.~\ref{fig:circuit}(b), the current of density $\beta{J}/2$ flows from the spin-down to the spin-up channel in the left half of the spin valve, which is similar to the AP configuration. However, the current of density $\beta{J}/2$ flows inversely in the right half because the sign of the spin accumulation and associated spin relaxation is switched in this half. Similarly, the current density $J_\mathrm{sf,F}^\mathrm{P}=\alpha_\mathrm{F}^\mathrm{P}J/2$ and $J_\mathrm{sf,N}^\mathrm{P}=\alpha_\mathrm{N}^\mathrm{P}J/2$ can be derived from the circuit by using $J_\mathrm{sf,F}^\mathrm{P}+J_\mathrm{sf,N}^\mathrm{P}=\beta{J}/2$. One can also recover Eqs.~\eqref{totalh1}, \eqref{totalh2}, \eqref{sigmanpsc}, and \eqref{sigmanpsf} using Joule's law.

In summary, our analytical results show that the spin-dependent heat generation arises from two mechanisms: the spin relaxation and the spin diffusion. In a typical spin valve, the two mechanisms have equal contributions in the semi-infinite FM layers. However, in the NM spacer layer of a thickness much shorter than its spin-diffusion length, the spin-dependent heat generation is dominated by the spin relaxation in the AP configuration, and by the spin diffusion in the P configuration. The heat generation due to the two mechanisms can be expressed formally as the Joule heat of effective resistances.

\begin{acknowledgments}
We thank Prof.~Y. Suzuki and Prof.~A. A. Tulapurkar for fruitful discussions. This work was supported by National Natural Science Foundation of China (grant numbers~11404013, 11605003, 61405003, 11174020, 11474012) and 2016 Graduate Research Program of BTBU.
% 1.Yao-Hui Zhu; 2. Jian Liu; 3. Pei-Song He; 4. Bao-He Li; 5. Jun-Jie Shi
\end{acknowledgments}

\bibliography{heatbibfile}

\end{document}